
\input harvmac
\def\As{{\rm Ashtekar}}

\def\ut#1{\rlap{\lower1ex\hbox{$\sim$}}#1{}}
\def\sut#1{\rlap{\lower2ex\hbox{$\,\tilde{}$}}#1{}}

\def\eps{\epsilon}

\def\cW{{\cal W}}

\def\ha{{1\over 2}}
\def\w{\wedge}
\def\Newt{{1\over 16\pi G}}
\def\Ttitle#1#2#3{\nopagenumbers\abstractfont\hsize=\hstitle\rightline{#1}%
\nopagenumbers\abstractfont\hsize=\hstitle\rightline{#2}%
\vskip 1in\centerline{\titlefont #3}\abstractfont\vskip .5in\pageno=0}

\lref\wtqft{E.\ Witten, Comm.\ Math.\ Phys. {\bf 117}, (1988)353.}
\lref\tqft{This subject is reviewed in D.\ Birmingham, M.\ Blau, M.\ Rakowski,
and G.\ Thompson, Phys.\ Rep. {\bf 209}, (1991)129.}
\lref\don{S.\ K.\ Donaldson, Topology {\bf 29}, (1990)257.}
\lref\ash{A.\ Ashtekar, Phys.\ Rev.\ Letts. {\bf 57}, 2244(1986);
Phys.\ Rev. {\bf D36}, 1587(1986);
{\it New perspectives in canonical gravity}, (Bibliopolis, Naples,
1988);  {\it Lectures on non-perturbative canonical gravity},
(World Scientific, Singapore, 1991).
}
\lref\samuel{J.\ Samuel, Pram$\bar{\rm a}$na J.\ Phys. {\bf 28}, (1987)L429;
Class.\ Quant.\ Grav. {\bf 5}, (1988)L123.}
\lref\adm{R.\ Arnowitt, S.\ Deser, and C.\ W.\ Misner, Phys.Rev.
{\bf 120}, (1960)313; Ann.Phys. {\bf 33}, (1965)33.}
\lref\cs{L.\ N.\ Chang and C.\ P.\ Soo, VPI-IHEP 92-5,
{\it Classical and quantum gravity in Ashtekar variables.}}
\lref\kod{H.\ Kodama, Phys.Rev. {\bf D42}, (1990)2548.}
\lref\csd{L.\ N.\ Chang and C.\ P.\ Soo, VPI-IHEP 91-2,
{\it Ashtekar's variables and the topological phase of quantum gravity},
Proceedings of the XXth.\ Conference on Differential Geometric Methods
in Physics, ed.\ S.\ Catto, A.\ Rocha, (World Scientific, Singapore,
1991.)}
\lref\ashren{A.\ \As\ and P.\ Renteln, {\it Lecture notes on new
variables,} Astrophys.\ Grp., Math.Dept., University of Poona (1987).}
\lref\wsigma{E.\ Witten, Comm.Math.Phys. {\bf 118}, (1988)411.}
\lref\polya{A.\ M.\ Polyakov, Mod.Phys.Lett. {\bf A11}, (1987)801.}
\lref\donbook{For a review, see S.\ K.\ Donaldson and P.\ B.\ Kronheimer,
{\it The geometry of four-manifolds,} (Oxford Science Publications,
Clarendon Press, Oxford, 1990).}
\lref\wthree{E.\ Witten, Nucl.Phys. {\bf B311}, (1988)46.}

\Ttitle{VPI-IHEP-92-4}{hepth@xxx/9203014}{BRST Cohomology and Invariants of
4D Gravity}
\vskip -0.20in
\centerline{{\titlefont in Ashtekar Variables}}
\vskip 0.5in
\centerline{Lay Nam Chang\foot{e-mail:laynam@vtvm2.bitnet} 
\& Chopin Soo\foot{e-mail:soocp@vtcc1.bitnet}}
\centerline{Institute for High Energy Physics}
\centerline{Virginia Polytechnic Institute and State University}
\centerline{Blacksburg, Virginia 24061-0435}
\bigskip\bigskip

\centerline{{\bf Abstract}}

We discuss the BRST cohomologies of the invariants associated with the
description of classical and quantum gravity in four dimensions,  using
the \As\ variables.  These invariants are constructed from several BRST
cohomology sequences.  They provide a systematic and clear characterization
of non-local observables in general relativity with unbroken diffeomorphism
invariance, and could yield further differential invariants for four-manifolds.
The description includes fluctuations of the vierbein fields, but there exits
a non-trivial phase which can be expressed in terms of Witten's topological
quantum field theory.  In this phase, the descent sequences are degenerate,
and the corresponding classical solutions can be identified with the
conformally self-dual sector of Einstein manifolds.   The full
theory includes fluctuations which bring the system out of this sector
while preserving diffeomorphism invariance.

\Date{3/92}

A great deal has been learned recently about 2-manifolds, and it is now
possible to give a complete description
in terms of the gravitational fields in two dimensions (2D)\polya.
Similar results have also been obtained in three dimensions\wthree.
It is however not entirely clear how
such results can be extended to 4-manifolds.   There are several reasons
for suspecting that this generalization will not be straightforward.
Firstly, unlike in two and three dimensions, in four
dimensions
pure gravity can
exist in a phase wherein
there are two propagating degrees of freedom.  Secondly, the work of
Donaldson and others\donbook\ shows that 4-manifolds can have far
richer differential structures than in other dimensions, so their complete
characterization can prove to be a daunting task indeed.

Witten has suggested\wtqft\wsigma\ that 4D gravity possesses a phase
describable
by a topological quantum field theory(TQFT)\tqft,
in that the
observables consist entirely of global invariants.
Some of these invariants have been identified.
In particular,
it can be shown that by considering the moduli space of
(anti-) instantons, the Donaldson maps\don\ can be identified as BRST
invariants
of the corresponding TQFT\wtqft.

What happens when we have 4-manifolds which can support 4D gravity with
propagating degrees of freedom?   To extend Witten's analysis to such
instances, it will be necessary to describe these degrees of freedom using
variables that are related naturally to those employed in TQFT.   In
particular,
we must require that they are suitable for implementing both
diffeomorphism and gauge invariance.

In this article,  we show that in many respects the variables introduced by
\As\ satisfy these requirements\ash.  In place of the
metric of general relativity, the classical \As\ variables corresponding to
specific Einstein manifolds
consist of densitized
triad fields, and their conjugate momenta, which turn out to be
simply $SO(3)$ gauge potentials satisfying (anti-) self-duality conditions.
We shall analyze the classical and BRST symmetry of 4D gravity in terms
of these variables, determine their cohomology descent equations, and
identify the invariants which can be constructed out of them.  These
quantities can then be used to describe the diffeomorphism and gauge invariant
observables of the theory, and to furnish characterizations
of the differential structures of such 4-manifolds.


We shall work with Riemannian manifolds and start with the action
proposed by Samuel\samuel:
\eqn\action{
{\cal A} ={1\over {16\pi G}}\int_{M}(2F_a\w \Sigma_a+ {\lambda\over
3}\Sigma_a\w \Sigma_a)
}
where the anti self-dual two-form
\eqn\anti{
\Sigma_a=\ha\epsilon_a\,^{bc} \left\{e_b\w e_c-e_0\w e_a\right\}
}
and $e_A,A=0,\ldots,3$ denote the vierbein one-forms in four dimensions.
The Latin indices run from 1 to 3 and label
internal $SO(3)$ indices. $F_a$ is the curvature two-form of the $SO(3)$
Ashtekar connection $A_a$, while $\lambda$ is the cosmological constant. In
the Ashtekar formalism, the metric is considered to be a derived
quantity, expressed in terms of $\Sigma$ through
\eqn\metric{
\tilde g_{\mu\nu}={1\over 12}
\epsilon_{abc}\tilde
\epsilon^{\alpha\beta\gamma\delta}\Sigma^a_{\alpha\beta}
\Sigma^b_{\gamma_\mu}\Sigma^c_{\nu\delta}
}

In applying the canonical formalism to Eqn.\ \action, it is
convenient to work in the spatial gauge in which the vierbein can be
written as
\eqn\vier{
e_{A\mu}=\left[\matrix{ N&0\cr N^je_{aj}& e_{ai} \cr}\right]
}
where the Greek index $\mu$ runs from 0 to 3. The form assumed in \vier\
is compatible with the ADM\adm\ decomposition of the metric
\eqn\adm{\eqalign{
ds^{2} =& e_{A\mu}e_{A\nu} dx^\mu dx^\nu \cr
=& N^2(dx^0)^2+g_{ij}(dx^i+N^idx^0)(dx^j+N^jdx^0) \cr
}}
with the spatial metric $g_{ij}=e^{a}\,_{i}e_{aj}$. Thus we see that the
choice \vier\ in no way compromises the values of the lapse and shift
functions $N$ and $N^i$ which have geometrical interpretations in
hypersurface deformations. With this decomposition, it is
straightforward to re-write Eqn.\ \action\ as
\eqn\raction{\eqalign{
{\cal A} =&%
\Newt\int d^4x \left\{2\tilde\sigma^{ia}\dot A_{ia}+ 2A_{0a}%
D_i\tilde\sigma^{ia}+ 2N^j\tilde\sigma^{ia}
F_{ija}\right\} \cr
   ~~~ &\qquad -\Newt\int d^4x \left\{\ut{N}\left(\epsilon_{abc}
\tilde\sigma^{ia}\tilde\sigma^{jb}F^c_{ij}+{\lambda\over
3}\epsilon_{abc}\sut{\eps}
_{ijk}\tilde\sigma^{ia}\tilde\sigma^{jb}\tilde\sigma^{kc}\right)\right\} \cr
   ~~~\qquad\qquad\qquad  & \qquad +{\rm boundary~ terms } \cr
}}
with $\tilde\sigma$ and $\ut{N}$
defined as follows
\eqna\density
$$\eqalignno{
\tilde\sigma^{ia}\equiv& \ha \tilde
\epsilon^{ijk}\epsilon^{abc}e_{jb}e_{kc}  &\density a\cr
\ut{N}\equiv& \det (e_{ai})^{-1}N  &\density b\cr
}$$
The tildes above and below the variables indicate that they are tensor
densities of weight 1 and -1 respectively. Thus $2\tilde\sigma^{ia}$ is
readily identified as the conjugate variable to $A_{ia}$. (We shall
suppress the factor $16\pi G$ for convenience). The variables $A_{0a},\
N^i$ and $\ut{N}$ are clearly Lagrange multipliers for
the Ashtekar constraints.
These constraints can be identified as
Gauss' law generating $SO(3)$ gauge invariance
\eqn\gauss{
G^a\equiv 2\ D_i\tilde\sigma^{ia}\approx 0
}
and the ``supermomentum'' and ``superhamiltonian'' constraints
\eqna\super
$$       \eqalignno{
H_i &\equiv 2\tilde\sigma^{ja}F_{ija}\approx 0 &\super a\cr
H&\equiv
\epsilon_{abc}\tilde\sigma^{ia}\tilde\sigma^{jb}(F^c_{ij}+{\lambda\over
3}\epsilon_{ijk}\tilde\sigma^{kc})\approx 0 &\super b\cr
}$$
Ashtekar showed that these constraints, despite their remarkable
simplicity, are equivalent in content to the constraints and constraint
algebra of $4D$ general relativity. The equations of motion that are
obtained can be written succinctly as\cs\
\eqna\eom
$$\eqalignno{
D\Sigma_a&= 0 &\eom a\cr
F_a&= S_{ab}\Sigma^b &\eom b\cr
}$$
with
$$
S_{ab}=S_{ba} \qquad \hbox{and} \qquad  \Tr S =-\lambda  \eqno(10c)
$$
The last two conditions solve the ``supermomentum'' and
``superhamiltonian'' constraints, and can be taken to be the general
solution for non-degenerate
metrics.
For metrics of Euclidean signature, $S$ is a real symmetric matrix with
three eigenvectors.  The classical solutions can then be classified according
to the number of distinct eigenvalues of $S$, and are called Types
I, D, or O, depending on there being three, two, or one, distinct
eigenvalues respectively.
The set of equations \eom\ can be shown to be
equivalent to the equations for Einstein manifolds in four
dimensions i.e. $R_{\mu\nu}=\lambda g_{\mu\nu}$\foot{%
We show elsewhere\cs\ that all classical Einstein
manifolds can be described as anti-instantons in the Ashtekar variables.
Indeed, we can establish a relationship between the reduced phase space
of these variables and the moduli space of reducible and irreducible
anti-instantons, and thence construct new topologically non-trivial
solutions.}.
One can, in principle,
eliminate $S$ from the theory by using the relation implied in \eom{b}
and obtain
\eqn\s{
S_{ab}=-\ha \ast(F_a\w\Sigma_b)
}
with
\eqn\eb{
\epsilon_a\,^{bc}F_b\w\Sigma_c=0
}
and
\eqn\fs{
F_a\w\Sigma_a=-2\lambda\ast(1)
}
where $\ast$ is the hodge dual operator.  $S$ plays an important
role in
characterizing the possible phases of the theory.
However, it is prudent not to
eliminate $\Sigma$ in favor of $F$ by inverting $S$, because
there are various interesting cases
for which $S$ is
non-invertible, and yet the vierbein and $\Sigma$  remain regular.
Explicit examples, especially those involving the $F=0$ sector, as well as
cases with
abelian anti-instantons,
are described in \cs.
In what follows,  we shall
construct invariants that are dependent on $F$ and on $\Sigma$,
thereby demonstrating that they are both essential in the description
of 4D gravity.

We begin by analyzing the symmetries of the
action and the associated BRST invariance.
It is easy to see that the action is invariant under $SO(3)$
 gauge
transformations as well as four dimensional diffeomorphisms.
 Working with ${\cal A}$, which is explicitly gauge-invariant,
we can consider a diffeomorphism $\phi$, of $M$ into itself, generated by
the
vector field $\beta$. On the one-form variables $A_a$ and $e_A$,
the
induced variations are Lie derivatives
\eqn\lie{
\delta A_a={\cal L}_\beta A_a=(i_\beta d+di_\beta)A_a
}
and
\eqn\liev{
\delta e_A={\cal L}_\beta e_A
}
with $i_\beta$ denoting interior multiplication or contraction with the
vector field $\beta$, which in local coordinates can be written as
$\beta^\mu\partial_\mu$. If we denote the Lagrangian four-form as $L$,
then
\eqn\invact{\eqalign{
\quad & \int_{M}(L-\phi^\ast L)= 0 \cr
\iff &\int_{M}{\cal L}_\beta L
= \int_{\partial M}i_\beta L = 0 \cr
}}
The action is thus invariant if $M$ is closed, while for
open $M$, invariance can be maintained provided $\beta$
vanishes at the boundary.
Since $SO(3)$ gauge invariance is a symmetry
of the
theory, one can also consider diffeomorphisms
for which the exterior derivative operator in
the Lie
 derivative is replaced by the
covariant derivative to make it
 compatible with the canonical analysis.

In the BRST formalism, the
classical gauge and diffeomorphism
symmetries are mimicked by transformations
with the parameters replaced by ghosts $\eta'_a$ and $\xi$.
 Thus the BRST transformations of the variables
are (henceforth $\delta$ shall mean $\delta_{BRST})$
\eqna\transf
$$\eqalignno{
\delta A_a&=-D\eta'_a+{\cal L}_\xi A_a &\transf a\cr
\delta e_a&=-\eps_{a}\,^{bc}\eta'_b e_c +{\cal L}_\xi e_a &\transf b\cr
\delta e_0&={\cal L}_\xi e_0  &\transf c\cr
}
$$
For a general differential form  $\chi$,
\eqn\liec{
{\cal L}_\xi\chi=(i_\xi d-di_\xi)\chi
}
There is a sign difference in the second term
 because, unlike a normal vector field, $\xi$ carries a ghost number of
1.
The above BRST transformation of the vierbein
implies that the antiself-dual two form $\Sigma_a$ transforms
according to
\eqn\sigtrnf{
\delta\Sigma_a=-\eps_{a}\,^{bc}\eta'_b\Sigma_c+{\cal L}_\xi\Sigma_a
}
The standard procedure of splitting the ghost $\eta'$ into
$\eta'=\eta-i_\xi A$
allows us to write the BRST transformations in a gauge-covariant
manner
\eqna\covtrnf
$$\eqalignno{
\delta A_a&=-D\eta _a+i_\xi F_a &\covtrnf a\cr
\delta\Sigma_a&=-\eps_{a}\,^{bc}\eta_b\Sigma_c+i_\xi
D\Sigma_a-Di_\xi\Sigma _a  &\covtrnf b\cr
}$$
The BRST transformations for the ghosts are
\eqn\ghtrnf{
\delta \eta_a=-\ha \eps_{a}\,^{bc}\eta_b\eta_c+\ha i_\xi i_\xi
F_a
}
and
\eqn\ghdiff{
\delta\xi=\ha\cal L_\xi\xi\quad {\rm i.e. }\quad
\delta\xi^\mu=\xi^\nu\partial_\nu\xi^\mu
}
It can be verified that the BRST transformations above are
nilpotent i.e. $\delta^2=0$. The variables $(A,\eta,\Sigma, \xi)$, which
are $(1,0,2)$ forms and a vector-field respectively, are assigned ghost
numbers $(0,1,0,1)$ and carry a grading equal to the form degree plus
the ghost number. By this we mean that if $\chi_{1,2}$ are
$p_{1,2}$-forms with ghost numbers $g_{1,2}$, then
\eqn\ghwed{
\chi_1\w\chi_2=(-1)^p\chi_2\w\chi_1;\quad
p=(p_1+g_1)(p_2+g_2)
}

The effective quantum action will
consist of the classical action and
 a piece from a gauge-fixing Lagrangian of the form
$L_{g.f.}=\delta\chi$ which will involve anti-ghosts
and auxiliary
fields. Since the BRST transformation
is nilpotent, the effective action
will be BRST- invariant provided the boundary
term vanishes.


The transformation rules for $A$ and $\eta$ may also be obtained
by
considering the multiplet  $(A,\eta)$  as a
connection $\widetilde A$
of the universal bundle over $M\times {\cal C}/{\cal G}$.
Here ${\cal C}$ is the space of connections $A$, while ${\cal G}$ is the
group of gauge transformations.
$\widetilde A$
carries a grading
of
1, and can
be decomposed into its  $( 1, 0 )$ and $( 0, 1 )$  components
as
\eqn\pot{
\widetilde A_a=A_a+\eta_a
}
Its curvature is given by
\eqn\curv{
\widetilde F_a=(d+\delta)\tilde A_a+\ha\eps_{a}\,^{bc}\widetilde
A_b\w\widetilde A_c
}
The transformation rules for $A$ and $\eta$ are equivalent to the
statement
\eqn\soul{
\widetilde F_a=\exp (i_\xi)F_a=(I+i_\xi+{1\over {2!}}i_\xi i_\xi) F_a
}
which reduces to the ``soul-flatness'' condition, $\widetilde F=F$,
in ordinary
gauge theories when diffeomorphism  invariance generated by $\xi$ is
absent.  Here the concept of ``horizontality'' in curved space is the
statement that $\widetilde F$ can be expanded in terms of $F$
 and its contractions with the ghost $\xi$. The BRST transformation of
$F$ as
a consequence of \covtrnf{a}\ is
\eqn\ftrnf{
\delta F_a=-\eps_{a}\,^{bc}\eta_b F_c-(Di_\xi F)_a
}
It can be verified that the curvature $\widetilde F$ satisfies the Bianchi
identity
\eqn\bianchi{
\widetilde D \widetilde F_a=(d+\delta)\widetilde F_a+\eps_{a}\,^{bc}\widetilde
A_b\w\widetilde F_c=0
}
As a consequence $\Tr (\widetilde F^n)$ obeys
\eqn\char{
\widetilde D(\Tr (\widetilde F^n))=0
}
Since the gauge group is $SO(3)$, it suffices to consider $n=2$. One
can expand $\widetilde F_a\w\widetilde F_a$ in terms
of the ghost number
i.e. writing
\eqn\ghno{
\widetilde F_a\w\widetilde F_a=\sum^4_{g=0} W^g_{p=4-g}
}
The resulting BRST cohomology descent equations from \char\ i.e.
$$(d+\delta)(\widetilde F_a\w\widetilde F_a)=0$$
are
\eqna\cw
$$\eqalignno{
dW^0_4&=0 &\cw a\cr
\delta W^g_{4-g}&=-dW^{g+1}_{3-g}\qquad\qquad {\rm for~} g=1,2,3 &\cw
b\cr
\delta W^4_0&=0 &\cw c\cr
}$$
with
$$ \eqalign{
W^0_4&=F_a\w F_a \qquad\qquad \qquad\quad W^1_3=i_\xi(F_a\w F_a)\cr
W^2_2&={1\over {2!}} i_\xi i_\xi(F_a\w F_a)\qquad\qquad W^3_1={1\over %
{3!}} i_\xi i_\xi i_\xi(F_a\w F_a)\cr
W^4_0&={1\over{4!}} i_\xi i_\xi i_\xi i_\xi(F_a\w F_a)}
$$



Consequently, there is an off-shell descent sequence involving
the curvature of the Ashtekar connection. In view of the
 fact that on-shell, the covariant curl of $\Sigma$ is zero
and $\Sigma$  transforms in the same way as $F$, one expects
that a BRST cohomology sequence
analogous to the \cW\ descent exists for $\Sigma$ as well.
This is indeed true
if one
also makes use of the other equations of motion  \eom{b}\ and \eom{c}. This
 suggests that even off-shell, a descent involving $\Sigma$ could be
 realized. However for off-shell computations we should
 keep all terms involving $D\Sigma$. We thus find that the BRST
transformations of $\Sigma$  and $\xi$, Eqns.\ \covtrnf{}\ and
\ghtrnf\ imply
\eqna\useq
$$\eqalignno{
\delta(i_\xi\Sigma_a)=&-\eps_{a}\,^{bc}\eta_b i_\xi\Sigma_c+ {1\over{2!}}
i_\xi i_\xi D\Sigma_a-{1\over{2!}}[D(i_\xi i_\xi \Sigma)]_a &\useq a\cr
\delta({1\over{2!}} i_\xi i_\xi \Sigma_a)=& -\eps_{a}\,^{bc}\eta_b{1\over
{2!}} i_\xi i_\xi\Sigma_c+ {1\over{3!}} i_\xi i_\xi i_\xi D\Sigma _a %
&\useq b\cr
}$$
So instead of the Bianchi identity for $\widetilde F$, we have
\eqna\ubianchi
$$\eqalignno{
\widetilde D\widetilde\Sigma_a =&(d+\delta)(\Sigma_a+i_\xi\Sigma_a+
{1\over{2!}}
i_\xi
i_\xi\Sigma_a) \cr
{}~~~ &\qquad +\eps_{a}\,^{bc}(A_b+\eta_b)\w(\Sigma_c+i_\xi\Sigma_c+ %
{1\over{2!}} i_\xi i_\xi\Sigma_c)  \cr
{}~~~=&D\Sigma_a + i_\xi D\Sigma_a + {1\over{2!}} i_\xi i_\xi D\Sigma_a + %
{1\over{3!}} i_\xi i_\xi i_\xi D\Sigma_a    &\ubianchi a\cr
}$$
i.e.
$$
\widetilde D\widetilde\Sigma_a=\exp (i_\xi)D\Sigma_a  \eqno(33b)
$$
Thus we see that the consistency condition is
the requirement that $\widetilde D\widetilde\Sigma$ can be expanded
in terms of $D\Sigma$ and
its contractions with the ghost $\xi$. For $\widetilde D\widetilde F$, this
expansion
is trivial because of the Bianchi identity for $F$. By considering
$\widetilde D(\widetilde\Sigma_a\w\widetilde\Sigma_a)$ we have
\eqn\dd{
(d+\delta)(\widetilde\Sigma_a\w\widetilde\Sigma_a)=2\widetilde
D\widetilde\Sigma_a\w\widetilde\Sigma_a
}
and expanding $\widetilde\Sigma_a\w\widetilde\Sigma_a$ in terms of ghost
number,
$$\widetilde \Sigma_a\w\widetilde\Sigma_a=\sum^4_{g=0} V^g_{4-g}
$$
and using  \covtrnf{b}, \useq{}\ and \ubianchi{},  Eqn.\ \dd\  yields the
identities
\eqna\udesc
$$\eqalignno{
d(\Sigma_a\w\Sigma_a)=& 2D\Sigma_a\w\Sigma_a  &\udesc a\cr
\delta(\Sigma_a\w\Sigma_a)=& -d[i_\xi
(\Sigma_a\w\Sigma_a)]+i_\xi d(\Sigma_a\w\Sigma_a)  &\udesc b\cr
\delta[i_\xi (\Sigma_a\w\Sigma_a)]=&-d[{1\over{2!}}
i_\xi i_\xi (\Sigma_a\w\Sigma_a)]+ {1\over {2!}} i_\xi i_\xi
 d(\Sigma_a\w\Sigma_a) &\udesc c\cr
\delta[{1\over{2!}}i_\xi i_\xi(\Sigma_a\w\Sigma_a)]=&-d[{1\over{3!}}
i_\xi i_\xi i_\xi(\Sigma_a\w\Sigma_a)]+ {1\over{3!}} i_\xi i_\xi
i_\xi d(\Sigma_a\w\Sigma_a) &\udesc d\cr
\delta[{1\over{3!}}i_\xi i_\xi i_\xi(\Sigma_a\w\Sigma_a)]=&-d[
{1\over{4!}}
i_\xi i_\xi i_\xi i_\xi (\Sigma_a\w\Sigma_a)]+{1\over{4!}} i_\xi
i_\xi i_\xi i_\xi d(\Sigma_a\w\Sigma_a) &\udesc d\cr
\delta[{1\over{4!}}i_\xi i_\xi i_\xi
i_\xi(\Sigma_a\w\Sigma_a)]=&-%
{1\over{5!}}
i_\xi i_\xi i_\xi i_\xi i_\xi d(\Sigma_a\w\Sigma_a) &\udesc  f\cr
}
$$
It is remarkable that although $\Sigma$, unlike $F$, has a non-zero
covariant curl
off-shell, the non-exact terms on the
R.H.S. of \udesc{b}\ - \udesc{f}\  actually vanish because they are all
contractions
with
$\xi$ of $d(\Sigma_a\w\Sigma_a)$,
which is {\it zero in
four dimensions}.  Hence, we do have another BRST
cohomology sequence with descent equations
\eqna\ucoho
$$ \eqalignno{
dV^0_4=& 0 &\ucoho  a\cr
\delta V^g_{4-g}=& -dV^{g+1}_{3-g}\qquad\qquad\qquad {\rm for}\quad
g=1,2,3, &\ucoho b\cr
\delta V^4_0=&0  &\ucoho c\cr
}$$
where
$$ \eqalign{
V^0_4&=\Sigma_a\w\Sigma_a\qquad\qquad\qquad\quad
V^1_3=i_\xi(\Sigma_a\w\Sigma_a)\cr
 V^2_2&= {1\over{2!}}i_\xi i_\xi
(\Sigma_a\w\Sigma_a)\qquad\qquad V^3_1= {1\over{3!}} i_\xi i_\xi
i_\xi (\Sigma_a\w\Sigma_a)\cr
V^4_0&= {1\over{4!}} i_\xi i_\xi i_\xi i_\xi
(\Sigma_a\w\Sigma_a)
}$$

The existence of the two descents above naturally
leads us to ask whether a further descent can be
constructed from $\widetilde\Sigma_a\w\widetilde F_a$ by considering
$\widetilde D(\widetilde\Sigma_a\w\widetilde F_a)$. It is straightforward to
verify that from
$$(d+\delta)\widetilde\Sigma_a\w\widetilde F_a)=\widetilde D
\widetilde\Sigma_a\w\widetilde F_a
$$
we arrive at
\eqna\vdesc
$$\eqalignno{
\delta(\Sigma_a\w F_a)&=-d[i_\xi
(\Sigma_a\w F_a)]+i_\xi d(\Sigma_a\w F_a) &\vdesc   a\cr
\delta[i_\xi (\Sigma_a\w F_a)]&=-d[ {1\over{2!}}
i_\xi i_\xi (\Sigma_a\w F_a)]+ {1\over{2!}} i_\xi i_\xi
 d(\Sigma_a\w F_a) &\vdesc b\cr
\delta[{1\over {2!}}i_\xi i_\xi(\Sigma_a\w F_a)]&=-d[%
{1\over{3!}}
i_\xi i_\xi i_\xi(\Sigma_a\w F_a)]+ {1\over{3!}} i_\xi i_\xi
i_\xi d(\Sigma_a\w F_a) &\vdesc  d\cr
\delta[{1\over{3!}}i_\xi i_\xi i_\xi(\Sigma_a\w
F_a)]&=-d[%
{1\over{4!}}
i_\xi i_\xi i_\xi i_\xi (\Sigma_a\w F_a)]+ {1\over{4!}} i_\xi
i_\xi i_\xi i_\xi d(\Sigma_a\w F_a) &\vdesc  d\cr
\delta[{1\over{4!}}i_\xi i_\xi i_\xi
i_\xi(\Sigma_a\w F_a)]&=-%
{1\over{5!}}
i_\xi i_\xi i_\xi i_\xi i_\xi  d (\Sigma_a\w F_a)   &\vdesc  e\cr
}$$
The non-exact terms on the R.H.S. are again contractions
 with the ghost $\xi$  of $d(\Sigma_a\w F_a)$,  which also
vanish
in four dimensions. There is thus a
third BRST cohomology descent
sequence. While the  two previous
descents involve either $F$
 or $\Sigma$  but not both together, this third set of descent
equations
involves both $\Sigma$ and $F$ i.e. the conjugate variables
$\widetilde\sigma$ and $ A$. The
corresponding descent equations can be written as
\eqna\vcoho
$$\eqalignno{
dU^0_4&= 0 &\vcoho a\cr
\delta U^g_{4-g}&=-dU^{g+1}_{3-g}\qquad\qquad {\rm for}~ g=1,2,3, &\vcoho
b\cr
\delta U^4_0&=0 &\vcoho c\cr
}$$
with
$$\eqalign{
U^0_4&=\Sigma_a\w F_a\qquad\qquad\qquad\quad U^1_3=i_\xi(\Sigma_a\w
F_a)\cr
U^2_2&= {1\over{2!}} i_\xi i_\xi (\Sigma_a\w F_a)\qquad\qquad
U^3_1={1\over{3!}} i_\xi i_\xi i_\xi (\Sigma_a\w F_a)\cr
U^4_0&={1\over{4!}} i_\xi i_\xi i_\xi i_\xi (\Sigma_a\w
F_a)}$$

 The elements of the BRST descents can be used to construct invariants
by integrating over the appropriate cycles.
Observables of the theory should be expressed in terms of these
invariants.
To obtain the invariants,
consider  $\gamma_p\in H_p(M);\
p=0,\ldots,4$.
Picking an element $Y^{4-p}_p$ of the $W,V$, or $U$
descent, we see that
$$\eqalign{
\delta\int_{\gamma_p} Y^{4-p}_p &=-\int_{\gamma_p} d Y^{5-p}_{p-1}\cr
&= -\int_{\partial\gamma_p} Y^{5-p}_{p-1}\cr
&=0 }$$
Moreover, the BRST cohomology class of
$Y^{4-p}_p(\gamma_p)\equiv\int_{\gamma_p}Y^{4-p}_{p}$
depends only on the homology class of $\gamma_p $ because
the descent
equations guarantee that for
$\gamma_p$, $\zeta_p=\gamma_p+\partial\omega_{p+1}\in H_p(M);\
p=0,\ldots,3$
$$\eqalign{
Y^{4-p}_p(\zeta_p)&=\int_{\gamma_p+\partial\omega_{p+1}} Y^{4-p}_p\cr
&=Y^{4-p}_p(\gamma_p)+\int _{\omega_{p+1}} dY^{4-p}_p\cr
&=Y^{4-p}_p(\gamma_p)-\delta\int _{\omega_{p+1}} Y^{3-p}_{p+1}  \cr
}$$

Finally, consider the following transformation:
\eqn\new{
\delta\int_M S^{ab}F_a\w F_b=-\int_{\partial M}
i_\xi(S^{ab}F_a\w F_b)
}
Provided the boundary term vanishes (which is automatic if $M$
has
no boundary), a further global invariant, $\int_M S^{ab}F_a\w F_b$,
will be present.
 In the classical context,
this is an  independent invariant only for Petrov Type I Einstein
manifolds\cs.
It is possible to construct a descent for which $S^{ab}F_a \w F_b$
is the zero ghost number four-form.  This can be achieved by using
the identity
\eqn\newbian{\eqalign{
\widetilde{D}\left(S^{ab}\widetilde{F}_a \w \widetilde{F}_b \right)
   =& \left(\widetilde{D}S^{ab}\right)\w \widetilde{F}_a \w \widetilde{F}_b \cr
{}~~~  \qquad =& \left[\left(DS\right)^{ab} + i_{\xi}\left(DS\right)^{ab}%
\right] \w \widetilde{F}_a \w \widetilde{F}_b  \cr
}}
with $S_{ab} = -{1\over 4}{\ast}\left(F_a \w \Sigma_b + \Sigma_a \w %
        F_b \right)$.   The descent equations take the form
\eqna\newdesc
$$\eqalignno{
   d\left(S^{ab}F_a \w F_b \right) =& 0  &\newdesc a\cr
\delta(S^{ab}F_a \w F_b )=& -d[i_\xi
(S^{ab}F_a \w F_b )]   &\newdesc b\cr
\delta[i_\xi (S^{ab}F_a\w F_b ]=& -d[{1\over{2!}}
i_\xi i_\xi (S^{ab}F_a \w F_b )]
 &\newdesc c\cr
\delta[{1\over{2!}}i_\xi i_\xi(S^{ab}F_a \w F_b )]=&-d[{1\over{3!}}
i_\xi i_\xi i_\xi(S^{ab}F_a\w F_b)]
 &\newdesc d\cr
\delta[{1\over{3!}}i_\xi i_\xi i_\xi(S^{ab}F_a \w F_b )]=&-d[
{1\over{4!}}
i_\xi i_\xi i_\xi i_\xi (S^{ab}F_a \w F_b )]
 &\newdesc e\cr
\delta[{1\over{4!}}i_\xi i_\xi i_\xi
i_\xi(S^{ab}F_a\w F_b)]=& 0
&\newdesc  f\cr
}
$$
 Unlike the previous descents, this set of equations explicitly
involves the duality operator $\ast$, and hence the inverse of the metric
$g_{\mu\nu}$, defined through Eqn.\ \metric.   Since degenerate metrics could
not be
ruled out in quantum fluctuations, and the descent involves complicated
products of non-commuting operators, it remains to be seen whether the descent
\newdesc\ survives regularization.   However, even in the classical context, it
would be interesting to investigate the interplay between degenerate
metrics and the invariants defined by \newdesc{}.

In TQFT, the form with zero ghost number in the descent equations,
called the top-form, can be regarded as the action before gauge fixing.
For our purposes, the action \action\   is also
a combination of top-forms
from two of the three sequences, and correspond to $U^{0}_{4}$, and
the cosmological(volume)
4-form $V^{0}_{4}$.  The last
top-form, $W^{0}_{4}$ can be added on to the action density, and
will give a term analogous to the $\theta$-term in QCD.

A Dirac quantization of the theory requiring that physical
quantum states be
annihilated by all the constraints has to face the problem of having a
consistent ordering of the quantum constraints.
We choose one specific scheme of ordering to make the
discussion concrete.
While such a procedure is not the final word
 on such a complicated theory, it is nevertheless
 instructive, and one expects the broad
features of the theory to be present for all schemes.
 An ordering with a formal closure of the quantum
constraint algebra exists and it can be shown\kod\csd\ that the resulting
physical states  selected by
\eqn\phys{
\widehat H|\Psi>=\widehat G^a|\Psi>=\widehat H_i|\Psi>=0
}
has a sector described by
\eqn\ann{
\widehat Q|\Psi>=0
}
where
\eqn\brstch{
\widehat Q={\lambda\over 3}\int_{M^{3}}d^3x\delta
A_{ia}\exp \left(-i{{3C}\over \lambda}\right) {\delta\over{2i\delta A_{ia}}}
\exp \left(i{{3C}\over\lambda}\right)
}
and $C$ denotes the Chern-Simons functional
\eqn\chern{
C=\int_{M^{3}}d^3x\epsilon^{ijk} \left(A_{ia}\partial_j A^a_k+
{1\over 3}\epsilon^{abc} A_{ia} A_{jb} A_{ck}\right)
}
Indeed in the reduced phase space analysis,
the restricted sector of the theory corresponding
to \ann\ is described by the Ashtekar-Renteln ansatz\ashren\
\eqn\ashr{
B^{ia}=-{\lambda\over 3}\tilde\sigma^{ia}
}
This ansatz is the set
 of initial data for conformally self-dual
Einstein manifolds and they are described by
\eqn\conf{
F_a=-{\lambda\over 3}\Sigma_a
}
$S_{ab}=-({\lambda / 3})\delta_{ab}$ is precisely
Type O  for this sector of the full theory.
A naive counting of the number of constraints
 tells us that if we are restricted to this TQFT phase,
 there are no local degrees of freedom. In this phase, the reduced
action is
\eqn\redact{
{\cal A}_{TQFT} =-{3\over {16\pi\lambda G}}\int_{M} F_a\w F_a
}
The resulting constraints are Eqn.\ \ashr\ and Gauss' law.
This set of constraints, however, is reducible since \ashr\
generates deformations of the gauge potential, and so if it
holds, so will Gauss' law. The BRST analysis of this TQFT
action has been performed
(see\tqft\ for details) and the BRST invariants from the descent
equations were successfully identified with
the Donaldson maps by Witten\wtqft. If we let $\delta_{TQFT}$ be the
BRST transformation, then the action is invariant under
\eqna\tinv
$$\eqalignno{
\delta_{TQFT} A_a&=-D\eta_a+\psi_a &\tinv a\cr
\delta_{TQFT} \eta_a &=-\ha\eps_{a}\,^{bc}\eta_b\eta_c+\phi_a &\tinv b\cr
\delta_{TQFT} \psi_a&=-\epsilon_a\,^{bc}\eta_b\psi_c-D\phi_a &\tinv c\cr
\delta_{TQFT} \phi_a&=-\epsilon_a\,^{bc}\eta_b\phi_c &\tinv d\cr
}$$
Comparing with Eqns.\ \covtrnf{}\ and \ghtrnf, this restricted (Type O) phase
 of the full theory can be identified with the TQFT of Eqn.\ \redact\
with Donaldson-Witten invariants, if we make the substitution
\eqn\donwitt{
\psi_a=i_\xi F_a\qquad\qquad {\rm and}\qquad\qquad\phi_a=
{1\over{2!}}i_\xi i_\xi F_a
}
In this sector, the role of $\Sigma$  has been eliminated in terms of
$F$. From
the vantage point of the BRST invariants
and the observables we have
constructed so far,  we see that in
 this phase the elements of the descents are
{\it degenerate} because
 of \conf.

Eqn.\ \phys\ can have solutions which are not annihilated by the
topological charge $Q$ of Eqns.\ \brstch\ - \chern.  When this happens, we are
no longer in the TQFT phase, and classically at least, we will have solutions
which are not of Type O.  Outside of this phase, the variables $\Sigma$
and $F$ are no longer proportional to each other,
and so there can be
full-fledged canonical degrees of freedom in the theory.  A naive counting
of the number of constraints minus the number of conjugate pairs shows
that we now have up to two unconstrained degrees of freedom.
These represent
local fluctuations in the vierbeins and the \As\ connections\cs.
However, it is
important to remember that we still maintain complete diffeomorphism
invariance,  so it is not clear these degrees of freedom can be
directly identified with the graviton\wtqft\wsigma.
In any case, the
BRST descent sequences are now
independent of each other, and the observables are characterized by
many more invariants.
Despite the local nature of the fluctuations in $\Sigma$ and $F$,
the observables described by the invariants are non-local, since
exact diffeomorphism symmetry remains unbroken.    The physical
implications of these observables when this invariance is broken, as well
as the important question of what happens
in the presence of matter couplings,
will be discussed elsewhere.

\bigskip\bigskip

\centerline{\bf Acknowledgments}

We thank  Marek Grabowski and Waichi Ogura for helpful discussions.

\listrefs

\bye